\documentclass[letter]{jpsj3} 
\usepackage{color}
\usepackage{ulem}

\begin{document}
\title{Synchrotron X-ray Diffraction Study of Structural Phase Transition in Ca$_{10}$(Ir$_4$As$_8$)(Fe$_{2-x}$Ir$_x$As$_2$)$_5$}

\author{
Naoyuki Katayama$^{1}$\thanks{E-mail address: katayama@mcr.nuap.nagoya-u.ac.jp},
Kento Sugawara$^{1}$,
Yuki Sugiyama$^{1}$,
Takafumi Higuchi$^{1}$,
Kazutaka Kudo$^{2}$,
Daisuke Mitsuoka$^{2}$,
Takashi Mizokawa$^{3}$,
Minoru Nohara$^{2}$, and Hiroshi Sawa$^{1}$
}

\inst{
$^{1}$Department of Applied Physics, Nagoya University, Nagoya 464-8603, Japan\\
$^{2}$Department of Physics, Okayama University, Okayama 700-8530, Japan\\
$^{3}$Department of Physics and Department of Complex Science and Engineering, University of Tokyo, Kashiwa, Chiba 277-8561, Japan}

\abst{
We report on the structural phase transition found in Ca$_{10}$(Ir$_4$As$_8$)(Fe$_{2-x}$Ir$_x$As$_2$)$_5$, which exhibits superconductivity at 16 K. The $c$-axis parameter is doubled below a structural transition temperature of approximately 100 K, while the tetragonal symmetry with the space group $P4/n$ (No. 85) is unchanged at all temperatures measured. Our synchrotron X-ray diffraction study clearly shows iridium ions at a non-coplanar position shift along the $z$-direction at the structural phase transition. We discuss the fact that iridium displacements affect superconductivity in Fe$_2$As$_2$ layers.}


\maketitle

Since the discovery of superconductivity in LaFeAsO$_{1-x}$F$_x$ \cite{rf:1}, the high-$T_c$ superconducting mechanism has been attributed to both the magnetic and structural properties of the material. All iron-based superconducting families identified so far consist of the same structural motif of Fe$_2$As$_2$ and spacer layers \cite{rf:2, rf:4, rf:3, rf:5, rf:6, rf:7, rf:8, rf:9, rf:10, rf:11,rf: 12, rf:13, rf:14, rf:15, rf:16, rf:17, rf:19, rf:20}. Therefore, the crystal structure of a material can be classified according to the spacer layer. Examples include (i) 1111-type structures with slabs of rare-earth oxides or alkali-earth fluorides with a fluorite-type structure \cite{rf:1, rf:7}, (ii) 111- and 122-type structures with alkali or alkali-earth ions \cite{rf:8, rf:6}, (iii) a 32522-type structure and its derivative with complex metal oxides with a perovskite-type structure or combinations of perovskite- and rocksalt-type structures \cite{rf:9, rf:10, rf:11,rf: 12, rf:13, rf:14, rf:15, rf:16}, (iv) a 112-type structure with arsenic zigzag chains \cite{rf:2, rf:4}, and (v) 10-4-8 and 10-3-8 phases as in Ca$_{10}$(Pt$_n$As$_8$)(Fe$_{2-x}$Pt$_x$As$_2$)$_5$ with $n$ = 3 and 4 and Ca$_{10}$(Ir$_4$As$_8$)(Fe$_{2-x}$Ir$_x$As$_2$)$_5$ \cite{rf:3, rf:5, rf:19, rf:20}. Generally, spacer layers are insulating in nature without any degrees of freedom. If there are degrees of freedom of spacer layers, such as charges, orbitals and lattices, we expect that their orderings or fluctuations will affect the superconductivity in Fe$_2$As$_2$ layers through intimate interlayer coupling, leading to a new controllable parameter.

\begin{figure}[t]
\begin{center}
\includegraphics[width=7.6cm]{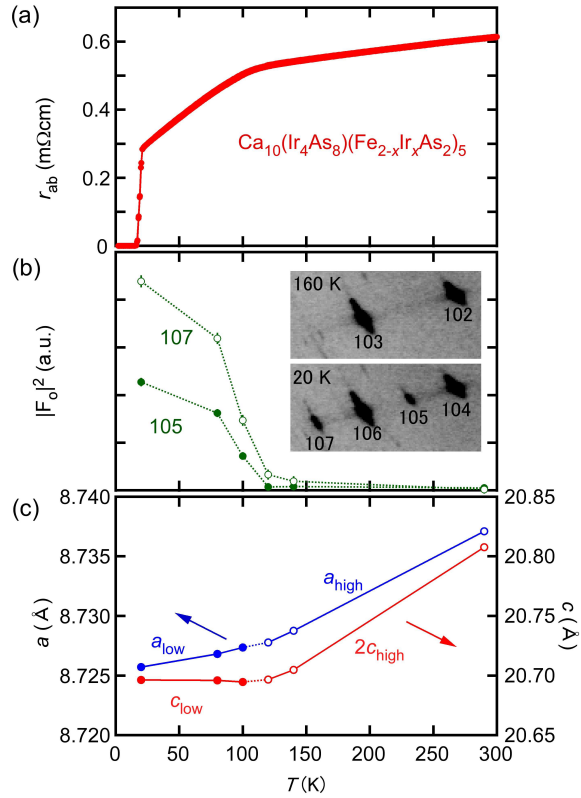}
\caption{\label{fig:Fig1}
(Color online)  (a) Temperature dependence of the electrical resistivity parallel to the $ab$-plane, $\rho_{\rm ab}$, for Ca$_{10}$(Ir$_4$As$_8$)(Fe$_{2-x}$Ir$_x$As$_2$)$_5$, which has been reported in Ref 15. Note that a small kink appears near 100 K. (b) Temperature dependence of the square of the observed structure factor, $|F_o|^2$, which is proportional to the superstructure peak intensity. (c) Lattice parameters at each temperature, where $c_{low}$ and $c_{high}$ indicate the $c$-axis parameters in the low-temperature and high-temperature phases, respectively. }
\end{center}
\end{figure}

To address the above mentioned issue, we present a structural study of Ca$_{10}$(Ir$_4$As$_8$)(Fe$_{2-x}$Ir$_x$As$_2$)$_5$ with a $T_c$ of 16 K. This compound was first reported by Kudo $et~al.$ with the chemical formula Ca$_{10}$(Ir$_4$As$_8$)(Fe$_2$As$_2$)$_5$ based on the synchrotron X-ray diffraction analysis \cite{rf:3}. In this study, we reanalyzed the chemical composition and found a small amount of Ir substitution for Fe, which will be discussed later. Both Ca$_{10}$(Ir$_4$As$_8$)(Fe$_{2-x}$Ir$_x$As$_2$)$_5$ and a platinum derivative, Ca$_{10}$(Pt$_4$As$_8$)(Fe$_{2-x}$Pt$_x$As$_2$)$_5$, reported by Ni $et~al.$, crystallize in tetragonal structures with the space group $P4/n$ \cite{rf:3, rf:5}. The most significant difference between them is related to their electron configurations: Pt$^{2+}$ forms a closed-shell configuration with a completely filled $d_{xy}$ orbital owing to the 5$d^8$ electric state, whereas Ir$^{2+}$ has a formally half-filled 5$d^7$ electric state. The two materials exhibit different temperature dependences in electrical conductivity, which is likely due to the different electron configurations of spacer layers. The resistivity of Ca$_{10}$(Pt$_4$As$_8$)(Fe$_{2-x}$Pt$_x$As$_2$)$_5$ decreases linearly with temperature \cite{rf:5}, whereas that of Ca$_{10}$(Ir$_4$As$_8$)(Fe$_{2-x}$Ir$_x$As$_2$)$_5$ exhibits a kink near 100 K, as shown in Fig. \ref{fig:Fig1}(a), suggesting that an unusual transition related to the electron configuration of the divalent iridium occurs. Kudo $et~al.$ suggested that the magnetic ordering is absent below 100 K based on the results of their M\"{o}ssbauer experiment \cite{rf:3}.

In this letter, we present a synchrotron X-ray structural study of Ca$_{10}$(Ir$_4$As$_8$)(Fe$_{2-x}$Ir$_x$As$_2$)$_5$. Our analysis clearly shows the appearance of superstructure peaks below 100 K, indicating a doubled period along the $c$-axis. In the low-temperature phase, half of the iridium ions shift along the $c$-axis, displacing the surrounding arsenic ions. 

\begin{table}[t]
\begin{center}
 \caption{Data collection and refinement statistics for the synchrotron X-ray structure determination of Ca$_{10}$(Ir$_4$As$_8$)(Fe$_{2-x}$Ir$_x$As$_2$)$_5$.}
 \label{tab:table1}
 \vspace{3mm}
{\tabcolsep=3mm
 \begin{tabular}{lll}
  \hline
  \hline
Ca$_{10}$(Ir$_4$As$_8$)(Fe$_{2-x}$Ir$_x$As$_2$)$_5$ & 293 K & 20 K\\
  \hline
  \hline
 \multicolumn{3}{c}{Data Collection }\\
  Crystal System & tetragonal & tetragonal \\
 Space Group & $P4/n$ & $P4/n$\\
 $a$ (\AA) & 8.73230(10) & 8.7236(17) \\
 $c$ (\AA) & 10.391(4) & 20.6799(18)  \\
 $R_{\rm merge}$ & 0.0613 & 0.0614\\
 $I$ / $\sigma$ $I$ & $>$2 & $>$2\\
 &\\
   \hline
 \multicolumn{3}{c}{Refinement }\\
 Resolution (\AA) &  $>$0.50 &  $>$0.50\\
 No. of Unique Reflections & 2329 & 3457\\
 $R$1  & 0.0369 & 0.0409\\
 \hline
  \hline
 \end{tabular}}
\end{center}
\end{table}

\begin{table}[hbt]
\renewcommand{\arraystretch}{1}
\caption{
Structural parameters collected at 293 K. $M$(3) and $M$(4) represent Fe$_{1-x/2}$Ir$_{x/2}$ with $x$ = 0.085(3) and 0.059(2), respectively. The occupancy is fixed at 1 in all atomic sites.}
\label{tab:table2}
\begin{tabular}{lccc}
\hline
\hline
\multicolumn{4}{c}{Ca$_{10}$(Ir$_4$As$_8$)($M_2$As$_2$)$_5$ at 293 K.}\\
\hline
\multicolumn{4}{c}{Atomic Position}\\
Site~~   & $x/a$ & $y/b$ & $z/c$ \\
\hline
Ir(1)~~  & 3/4 & 1/4 & -1/2 \\
Ir(2)~~  & 3/4 & 3/4 & -0.56465(4) \\ 
$M$(3)~~  & 3/4 & 1/4 & 0 \\
$M$(4)~~  & 0.05219(5) & 0.34938(6) & -0.00473(5) \\ 
As(5)~~  & 0.64044(4) & 0.50455(5) & -0.49958(4) \\ 
As(6)~~  & 1/4 & 1/4 & -0.14574(9) \\
As(7)~~  & 0.84898(4) & 0.45365(4) & -0.13567(4) \\ 
Ca(8)~~  & 0.55128(8) & 0.34551(9) & -0.26082(8) \\ 
Ca(9)~~  & 3/4 & 3/4 & -0.27298(17) \\
\hline
\hline
\end{tabular}
\end{table}

\begin{table}[hbt]
\renewcommand{\arraystretch}{1}
\caption{
Structural parameters collected at 20 K. Below the transition, all atomic sites, except for Ir(1), become two distinct sites. $M$(3a), $M$(3b), $M$(4a), and $M$(4b) represent  Fe$_{1-x/2}$Ir$_{x/2}$ with $x$ = 0.089(3) for $M$(3a) and $M$(3b), and $x$ = 0.062(2) for $M$(4a) and $M$(4b). Here, $M$(3a) and $M$(3b) are derived from $M$(3), while $M$(4a) and $M$(4b) are derived from $M$(4). The occupancy is fixed at 1 in all atomic sites. }
\label{tab:table3}
\begin{tabular}{lccccc}
\hline
\hline
\multicolumn{4}{c}{Ca$_{10}$(Ir$_4$As$_8$)($M_2$As$_2$)$_5$ at 20 K. }\\
\hline
\multicolumn{4}{c}{Atomic Position}\\
Site~~   & $x/a$ & $y/b$ & $z/c$ \\
\hline
Ir(1)~~  & 1/4 & -1/4 & 0.247974(14) \\
Ir(2a)~~  & -1/4 & -1/4 & 0.20958(2) \\ 
Ir(2b)~~  & 1/4 & 1/4 & 0.27449(2) \\ 
$M$(3a)~~  & -3/4 & -1/4 & 0 \\
$M$(3b)~~  & 1/4 & -1/4 & 1/2 \\
$M$(4a)~~  & -0.34874(9) & -0.44598(9) & 0.00378(4) \\ 
$M$(4b)~~  & -0.05035(9) & -0.35010(9) & 0.50078(4) \\
As(5a)~~  & -0.14200(8) & -0.00692(8) & 0.25072(3) \\ 
As(5b)~~  & 0.13935(8) & 0.00283(8) & 0.24954(3) \\
As(6a)~~  & -1/4 & -1/4 & 0.07647(6) \\
As(6b)~~  &  -1/4 & -1/4 & 0.56830(6) \\
As(7a)~~  & -0.54546(8) & -0.34881(8) & -0.06679(3) \\ 
As(7b)~~  & 0.04737(7) & -0.15073(7) & 0.43176(3) \\ 
Ca(8a)~~  & 0.05329(14) & -0.15489(14) & 0.12963(6) \\ 
Ca(8b)~~  & -0.04915(14) & 0.15412(15) & 0.36948(6) \\
Ca(9a)~~  & 1/4 & 1/4 & 0.13355(11) \\
Ca(9b)~~  & -1/4 & -1/4 & 0.36103(12) \\
\hline
\hline
\end{tabular}
\end{table}
%

Single crystals of Ca$_{10}$(Ir$_4$As$_8$)(Fe$_{2-x}$Ir$_x$As$_2$)$_5$ were grown at Okayama University. The synthesis details have already been reported \cite{rf:3}. The single-crystal X-ray diffraction experiments were performed at SPring-8 BL02B1 (Hyogo, Japan) using single crystals with typical dimensions of 40 $\times$ 40 $\times$ 70 $\mu$m$^3$. The X-ray wavelength was 0.52 \AA. The obtained lattice parameters and refined conditions are summarized in Table \ref{tab:table1}. 

\begin{figure*}[t]
\begin{center}
\includegraphics[width=14.2cm]{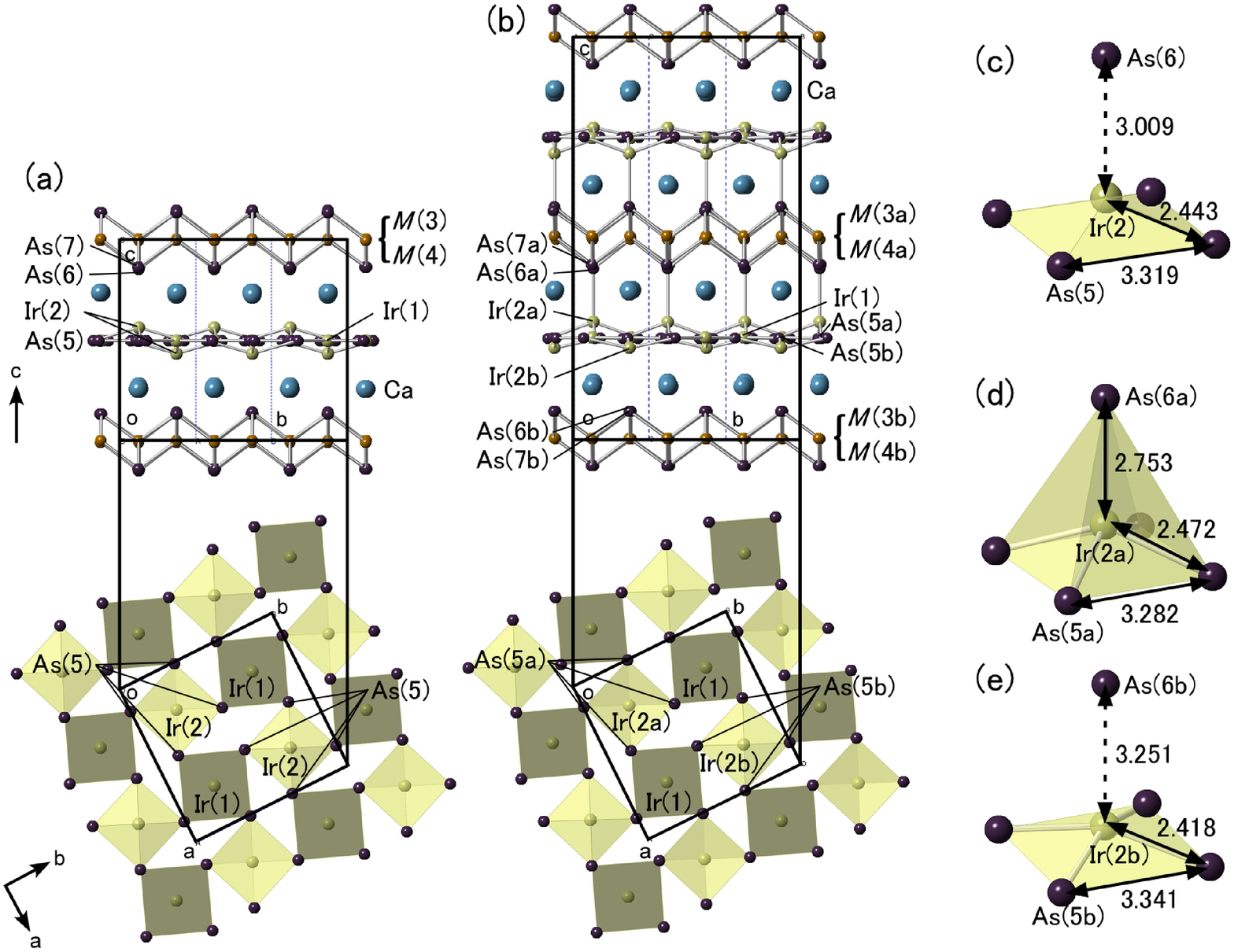}
\caption{\label{fig:Fig2}
(Color online) Crystal structures of Ca$_{10}$(Ir$_4$As$_8$)(Fe$_{2-x}$Ir$_x$As$_2$)$_5$ at (a) 293 and (b) 20 K. Atomic distances between the iridium and arsenic ligands, where (c) represents Ir(2) and the ligands at 293 K, while (d) and (e) show Ir(2a) and Ir(2b) with the arsenic ligands at 20 K, respectively. }    
\end{center}
\end{figure*}

In our single-crystal X-ray diffraction experiments, superstructure peaks emerged near 100 K, which nearly corresponded to the kink in the electric resistivity measurement, and the peaks increased in number as the sample was cooled, as shown in Fig. \ref{fig:Fig1}(b). The superstructure peaks indicate a doubled period along the $c$-axis direction below the transition. As shown in Fig. \ref{fig:Fig1}(c), the lattice parameters show no discontinuous jump at 100 K, suggesting that the transition is of the second order.

Through careful investigation of the Laue symmetry and the extinction rule of $h$+$k$$\neq$2$n$ for $hk$0, we found that the structural symmetry at 20 K was tetragonal with the space group $P4/n$, indicating that the space group was unchanged across the transition. Our precise definition of the composition showed a small amount of Ir substituted for Fe in the present samples, which was not reported in a previous study \cite{rf:3}. The high-temperature and low-temperature structural data are summarized in Tables \ref{tab:table2} and \ref{tab:table3}, respectively.  Crystal information files (CIFs) of the crystal structure of Ca$_{10}$(Ir$_4$As$_8$)(Fe$_{2-x}$Ir$_x$As$_2$)$_5$ at 293 and 20 K derived by analysis can be obtained free of charge from The Cambridge Crystallographic Data Centre via www.ccdc.cam.ac.uk. CCDC 1016394 and 1016395 contain the crystallographic data for this paper.

The space group $P4/n$ is preserved at the lowest measured temperatures, although some atomic sites become two distinct sites, as shown in Figs. \ref{fig:Fig2}(a) and \ref{fig:Fig2}(b). The most notable structural features appear near the iridium ions. There are two iridium sites at high temperatures: Ir(1) is at a coplanar site and Ir(2), which becomes two distinct sites of Ir(2a) and Ir(2b) at low temperatures, is at a non-coplanar site with respect to the As square. At the transition, Ir(2b) shifts toward the center of the As square, as shown in Figs. \ref{fig:Fig2}(c) and \ref{fig:Fig2}(e). In contrast, Ir(2a) is displaced toward the As(6a) ion in the adjacent Fe$_2$As$_2$ layer, resulting in an almost 10\% reduction in the Ir(2a)-As(6a) distance compared with that at 293 K, as shown in Figs. \ref{fig:Fig2}(c) and \ref{fig:Fig2}(d). Note that As(6a) also shifts toward Ir(2a) at low temperatures, suggesting a strong tendency toward the bonding between Ir(2a) and As(6a).

What is the origin of the unusual structural transition found using the synchrotron X-ray diffraction experiment? One of the possible scenarios relates to the band splitting of arsenic, which is analogous to the structural transition found in IrTe$_2$ with a two-dimensional triangular lattice \cite{rf:27}. IrTe$_2$ shows the structural transition caused by the crystal field effect, which further splits the energy levels of Te ($p_x$, $p_y$) and Te $p_z$. Such a tellurium-originated structural transition is absent in the platinum analogue PtTe$_2$, which also seems to be similar to the present systems. This scenario, however, seems not to be the case in the present systems because the displacements of Ir(2a) and Ir(2b) at the transition are much larger than those of the surrounding arsenic ions, which is sharply different from the structural transition in IrTe$_2$ with large displacements of tellurium. The experimentally observed large displacements of iridium ions make us expect the modification of bonding characters between iridium and surrounding arsenic ions.

Experimentally observed strong tendency toward bonding formation between Ir(2a) and As(6a) makes us expect that an unpaired electron occupies the Ir(2a) 5$d_{3z^2-r^2}$ orbital, which spreads toward As(6a) directions and contributes to the bonding between Ir(2a) and As(6a). Note that the divalent nature of iridium ions has already been confirmed by X-ray photoelectron spectroscopy (XPS) \cite{rf:3}, indicating the formal electron counts of 5$d^7$. To determine whether the unpaired electron indeed occupies 5$d_{3z^2-r^2}$ orbital, we performed the configuration interaction calculation on the IrAs$_5$ cluster model with $d^7$, $d^8L$, and $d^9L^2$ configurations, where $L$ represents a ligand hole. 

In the cluster model calculation, the ground and excited states are given by linear combinations of $d^7$, $d^8L$, and $d^9L^2$ configurations, where $L$ denotes a hole in the As 4$p$ orbitals. The charge-transfer energy $\Delta$ corresponds to $E$($d^8L$)$-E$($d^7$). Here, $E$($d^nL^m$) represents the average energy of the $d^nL^m$ states. $E$($d^9L^2$)$-E$($d^8L^2$) is given by $\Delta+U$, where $U$ is the multiplet average Coulomb interaction between the $d$ electrons. The Ir 5$d$ spin-orbit interaction $\zeta_d$ is set to 0.5 eV. The Racah parameters $B$ and $C$ for the Ir 5$d$ orbitals are set to zero since the Hund exchange energy ($5B/2+C$) is negligible in Ir 5$d$ compounds compared with $U$. The hybridization terms between $d^7$ and $d^8L$ (or $d^8L$ and $d^9L^2$) are given by ($pd\sigma$) and ($pd\pi$), where ($pd\sigma$)/($pd\pi$)=-2.16. $\Delta$, $U$,and ($pd\sigma$) are set to 2.0, 1.0, and -1.5 eV, respectively, considering the results of band calculations \cite{rf:3}, photoemission experiments, and chemical trends of the parameters. The present conclusion does not change if the three parameters are changed within a reasonable range. For example, the calculated results are essentially the same when ($pd\sigma$) is increased to -2.0 eV or decreased to -1.0 eV. Figure 3 shows the energies of the ground and excited states as a function of the Ir(2)-As(5)-As(6) bond angle $\theta$. The Ir(2b) site can be regarded as a square planar coordination since the apical As(6b) is far from the basal plane. As expected, the unpaired electron occupies 5$d_{3z^2-r^2}$ in the ground state. However, the energy splitting between 5$d_{3z^2-r^2}$ and 5$d_{xy}$ is as small as $\sim$0.2 eV, which is comparable to or smaller than the hybridization energy between neighboring clusters. Therefore, in the Ir(2b) site, the unpaired electron can be a mixture of 5$d_{3z^2-r^2}$ and 5$d_{xy}$. On the other hand, in the Ir(2a) site, the apical As(6a) approaches to the basal plane and, consequently, the unpaired electron tends to be pure 5$d_{3z^2-r^2}$ with a larger energy splitting. Therefore, the present cluster model calculation suggests that the structural distortion discovered in the present work is accompanied by the Ir 5$d$ orbital change. If the apical As(6a) is much closer to the basal plane ($\theta <$ 42$^\circ$), the unpaired electron is predicted to occupy 5$d_{x^2-y^2}$ in the ground state. Here, it should be noted that the impact of the Ir 5$d$ spin-orbit interaction on the Ir 5$d_{xy}$/5$d_{yz}$ orbitals is much stronger than those on the Ir 5$d_{3z^2-r^2}$, 5$d_{x^2-y^2}$, and 5$d_{xy}$ orbitals.

Although the displacement of Ir(2a) toward the As(6a) direction causes the contraction of the As(5a) square, as shown in Figs. \ref{fig:Fig2}(c) and \ref{fig:Fig2}(d), the As(5b) square expands below the transition temperature, accompanied by the displacement of Ir(2b) toward the center of the As square, as shown in Figs. \ref{fig:Fig2}(c) and \ref{fig:Fig2}(e), resulting in the appearance of the breathing square lattice in the Ir$_4$As$_8$ layer. Note that similar breathing lattices can also be found in other systems. The spinel oxides, LiGaCr$_4$O$_8$ and LiInCr$_4$O$_8$\cite{rf:22}, consist of an inherently distorted breathing pyrochlore lattice owing to the large size mismatch between Li$^+$ and Ga$^{3+}$/In$^{3+}$ ions at the $A$ site. The perovskite oxides CaFeO$_3$ \cite{rf:23}, YNiO$_3$ \cite{rf:24}, and NdNiO$_3$\cite{rf:25, rf:26}, belong to another system with a breathing lattice. The breathing distortion is realized in relation to the charge disproportionation of $B$ site transition metals. In the present system, any degrees of freedom of iridium should be related to the emergence of a breathing lattice, as presented above. Furthermore, the intimate coupling between Ir$_4$As$_8$ and the adjusting Fe$_2$As$_2$ layers also holds the key for realizing a breathing lattice.  If both Ir(2a) and Ir(2b) displace toward the same direction with regard to the As squares, both As(1a) and As(1b) squares should expand or contract together in relation with the displacement of iridium, leading to the destruction of the commensurate periodicity between Fe$_2$As$_2$ and Ir$_4$As$_8$ layers by an abrupt change in the size of the Ir$_4$As$_8$ planes across the transition. In the present system,  the commensurate periodicity is preserved between Fe$_2$As$_2$ and Ir$_4$As$_8$ layers without any discontinuous jumps in the $a$-axis parameter, which softens the change in the size of the Ir$_4$As$_8$ planes across the transition.

Finally, we discuss superconductivity at low temperatures. The present structural transition should affect superconductivity in Fe$_2$As$_2$ layers through intimate interlayer couplings. Associated with the iridium displacements, Fe$_2$As$_2$ layers become two crystallographically inequivalent Fe$_2$As$_2$ layers below the transition, as shown in Fig. \ref{fig:Fig2}(b): one layer includes As(6a), which is strongly coupled with Ir$_4$As$_8$ layers through Ir(2a)-As(6a) bondings, while the other layer contains As(6b), which is apart from the adjacent Ir$_4$As$_8$ layers. Therefore, the displacement of arsenic ions associated with iridium displacements introduces notable differences in the As-Fe-As bond angles between the two Fe$_2$As$_2$ layers; the angle ranges from 105.58(4) to 112.29(4)$^{\circ}$ for the layers containing As(6a) and from 107.14(4) to 110.69(4)$^{\circ}$ for the layers containing As(6b), which can be compared with the angle (109.47$^{\circ}$) of a regular tetrahedron. This indicates that any degrees of freedom of iridium composed of spacer layers can cause the divergence in the crystal field between two distinct Fe$_2$As$_2$ layers and the resulting rearrangement of energy bands around the Fermi level, which should significantly affect superconductivity.

Unlike Ca$_{10}$(Ir$_4$As$_8$)(Fe$_{2-x}$Ir$_x$As$_2$)$_5$, the platinum derivative Ca$_{10}$(Pt$_4$As$_8$)(Fe$_{2-x}$Pt$_x$As$_2$)$_5$ with space group $P4/n$ exhibits no structural transitions. Because Ca$_{10}$(Pt$_4$As$_8$)(Fe$_{2-x}$Pt$_x$As$_2$)$_5$ shows superconductivity with a higher transition temperature of 25 K \cite{rf:3, rf:5}, we can naively speculate that the structural transition suppresses superconductivity in the present system. However, we would like to point out that the second order structural transition in the present system may bring us further opportunities to study the relationship between superconductivity and structural instability, by suppressing the structural transition using any methods such as chemical doping or the application of pressure. 

\begin{figure}[thb]
\begin{center}
\includegraphics[width=8.2cm]{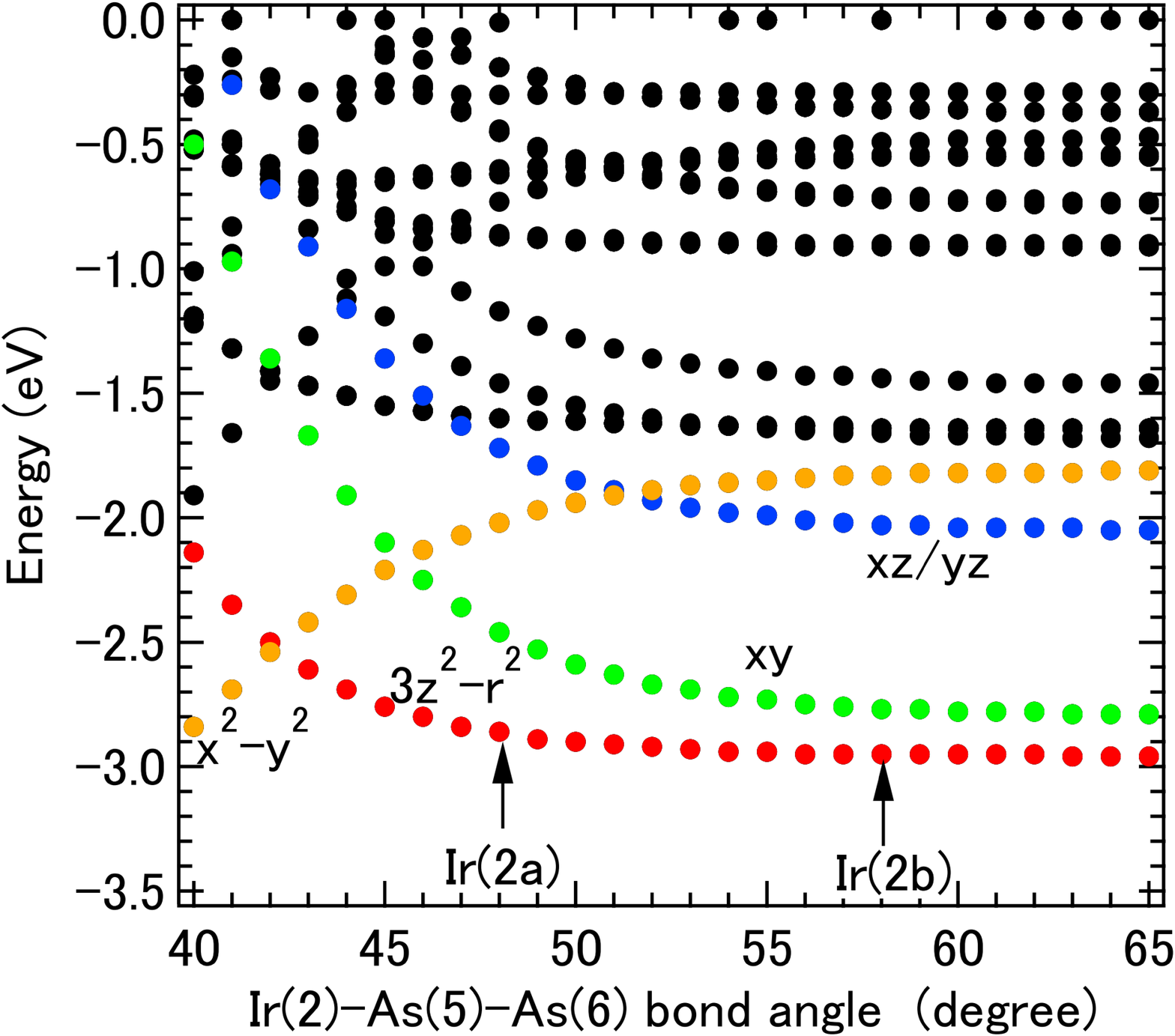}
\caption{\label{fig:Fig3}
(Color online) Energies of ground and excited states as a function of Ir(2)-As(5)-As(6) bond angles obtained from the IrAs$_5$ cluster model calculation.}
\end{center}
\end{figure}







In summary, we studied the nature of the structural transition found in Ca$_{10}$(Ir$_4$As$_8$)(Fe$_{2-x}$Ir$_x$As$_2$)$_5$, which undergoes a superconducting transition at 16 K, using synchrotron X-ray diffraction experiments. Our X-ray diffraction analysis results reveal the displacement of iridium and the associated arsenic displacement below the transition temperature. Combining these results with IrAs$_5$ cluster calculations, we conclude that all degrees of freedom of divalent iridium play an important role in structural transition. 

\begin{acknowledgements}
The work at Nagoya University was supported by a Grant-in-Aid for Scientific Research (No. 23244074), Kumagai Foundation for Science and Technology, Toukai Foundation for Technology, and Nippon Sheet Glass Foundation for Materials Science and Engineering. The work at Okayama University was partially supported by a Grant-in-Aid for Scientific Research (C) (No. 25400372) from the Japan Society for the Promotion of Science (JSPS) and the Funding Program for World-Leading Innovative R\&D on Science and Technology (FIRST Program) from JSPS. Some of the magnetization measurements were performed at the Advanced Science Research Center at Okayama University. The synchrotron radiation experiments performed at BL02B1 in SPring-8 were supported by the Japan Synchrotron Radiation Research Institute (JASRI) (Proposal Nos. 2012A0083, 2012B0083, 2013A0083, and 2013B0083).
\end{acknowledgements}

\end{document}